\documentclass{aa}

\usepackage[varg]{txfonts}
\usepackage{natbib,graphicx}

\bibpunct{(}{)}{;}{a}{}{,} 
\usepackage{color}

\begin{document}

\title{Magnetic field vector ambiguity resolution in a quiescent prominence observed on two consecutive days}

\author{T. Kalewicz \and V. Bommier}

\institute{LESIA, Observatoire de Paris, Universit\'e PSL, CNRS, Sorbonne Universit\'e, Univ. Paris Diderot, Sorbonne Paris Cit\'e 
\newline 5, Place Jules Janssen, 92190 Meudon, France}

\date{Received ... / Accepted ...}

\abstract
{Magnetic field vector measurements are always ambiguous, that is, two or more field vectors are solutions of the observed polarisation.}
{The aim of the present paper is to solve the ambiguity by comparing the ambiguous field vectors obtained in the same prominence observed on two consecutive days. The effect of the solar rotation is to modify the scattering angle of the prominence radiation, which modifies the symmetry of the ambiguous solutions. This method, which is a kind of tomography, was successfully applied in the past to the average magnetic field vector of 20 prominences observed at the Pic du Midi. The aim of the present paper is to apply this method to a prominence observed with spatial resolution at the TH\'EMIS telescope (European site at Iza\~na, Tenerife Island).}
{The magnetic field vector is measured by interpretation of the Hanle effect observed in the \ion{He}{i} D$_3$ 5875.6 \AA\, line, within the horizontal field vector hypothesis for simplicity. The ambiguity is first solved by comparing the two pairs of solutions obtained for a "big pixel" determined by averaging the observed Stokes parameters in a large region at the prominence centre. Each pixel is then disambiguated by selecting the closest solution in a propagation from the prominence centre to the prominence boundary.}
{The results previously obtained on averaged prominences are all recovered. The polarity is found to be inverse with a small angle of about $-21^{\circ}$ between the magnetic field vector and the long axis of the filament. The magnetic field strength of about 6 G is found to slightly increase with height, as previously observed. The new result is the observed decrease with height, of the absolute value of the angle between the magnetic field vector and the long axis of the filament.}
{This result is in excellent agreement with prominence magnetohydrodynamical models.}


\offprints{V. Bommier, \email{V.Bommier@obspm.fr}}

\titlerunning{Magnetic field vector ambiguity resolution in a quiescent prominence}
\authorrunning{T. Kalewicz \& V. Bommier}

\maketitle

\section{Introduction}

Following the pioneering work by \citet{Hyder-65} showing the possible existence of the Hanle effect in solar prominences, measurements of the effect were undertaken in order to diagnose the magnetic field in these objects. \citet{Leroy-77} \citep[see also][]{Leroy-etal-77} chose the \ion{He}{i} D$_3$ 5875.6 \AA\, line for this purpose because this line is absent from the incident photospheric spectrum, and therefore Doppler dimming or brightening is avoided. In addition, this line has the well-adapted Hanle sensitivity for determining the prominence magnetic field, which was found to be on the order of 6 G. Each spectral line has its Hanle sensitivity domain determined by the condition $\omega \tau \approx 1$, where $\omega$ is the Larmor pulsation depending on the field strength (1.4 MHz/G ignoring the Land\'e factor), and $\tau$ is the upper level radiative lifetime. A table of Hanle sensitivity for a series of spectral lines can be found in \citet{SahalB-81}, where it can be seen that the \ion{He}{i} D$_3$ line sensitivity fortunately lies around 6 G. 

Leroy measured the Hanle effect in about 400 quiescent prominences at the Pic du Midi coronagraph, during the ascending phase of solar cycle XXI (1976-1982), but without any spatial or spectral resolution in these objects.  Quiescent prominences are located far from any active region and/or are out of any eruptive phase. One single spectral line is insufficient to determine the three coordinates of the field vector, because the Hanle effect is described by two parameters only: the linear polarization degree and direction. Two lines of different Hanle sensitivity are required for this purpose. Fortunately, the \ion{He}{i} D$_3$ line (5875.6 \AA) is made of two partially overlapping components of different sensitivity, $3d^3D_{3,2,1} \rightarrow 2P^3P_{2,1}$ and $3d^3D_1 \rightarrow 2P^3P_0$, the latter of which is weaker and apart, but more polarisable. The peaks are separated by 0.43 \AA. Despite the difficulty in separating the components in the analysis \citep{Leger-Paletou-09,Koza-etal-17}, \citet{Athay-etal-83} analysed 13 quiescent prominences observed at Sacramento Peak with the Stokes-I \citep{Baur-etal-80} then Stokes-II \citep{Baur-etal-81} coronagraph spectropolarimeter, and clearly established the horizontality of the prominence magnetic field, with spatial resolution in prominences. \citet{Querfeld-etal-85} obtained similar results for two quiescent prominences also observed with Stokes-II. Leroy complemented his experiment with the quasi-simultaneous measurement of the Hydrogen H$\alpha$ and/or H$\beta$ line. From these data, \citet{Bommier-etal-86a} also obtained the horizontality of the field vector in quiescent prominences, and also a fourth parameter, the electron density \citep{Bommier-etal-86b}.

The theory of the Hanle effect was developed by \citet{Bommier-77}, \citet{Bommier-SahalB-78}, \citet{Bommier-80}, \citet{Landi-82}, and these first inversions were performed by linear interpolation in polarisation diagrams \citep{Bommier-etal-81} like those published by \citet{SahalB-etal-77}. 

Observation of the Hanle effect in solar prominences \citep{Schmieder-etal-13,Schmieder-etal-14a} were then performed at the French-Italian T\'elescope H\'eliographique pour l'\'Etude du Magn\'etisme et des Instabilit\'es Solaires (TH\'EMIS), installed at the European site at Iza\~na, in Tenerife. The \ion{He}{i} D$_3$ line was again observed, resolved in its two components. The horizontality of the field vector was recovered \citep{Schmieder-etal-14b}, and further investigations were possible inside the prominence fine structure with a better spatial resolution \citep{Levens-etal-16a,Levens-etal-16b,Levens-etal-17,Schmieder-etal-17}. The inversion method was based on the principal component analysis (PCA) technique \citep{LopezA-Casini-02,LopezA-Casini-03,Casini-etal-03,Casini-etal-05,Casini-etal-09}. The inversion code HAZEL for simultaneous analysis of Hanle and Zeeman effects was also developed \citep{AsensioR-etal-08}.

The analysis by \citet{Merenda-etal-06} alternatively found a far-from-horizontal magnetic field vector in a prominence observed by the Tenerife Infrared Polarimeter (TIP) mounted on the German Vacuum Tower Telescope (VTT) of the Observatorio del Teide (Tenerife, Spain). The analysed line is the polarisable component of the \ion{He}{i} 10,830 \AA\, line. For typical prominence magnetic fields, this line is not far from the saturated Hanle effect regime. \citet{Merenda-etal-06} came to the conclusion of a non-horizontal field from the fact that the observed linear polarisation degree and direction fall outside the horizontal field Hanle diagram, making a horizontal field incompatible with the observation. However, firstly the measurement inaccuracies and related box on the diagram are not reported, and secondly the observed polarisation is not far from the horizontal field diagram. Consequently, if the inaccuracy box had been taken into consideration, a horizontal field would have indeed been compatible with this observation. Moreover, the exact location of the filament on the surface of the  Sun remained undetermined, which was particularly difficult for this Polar Crown prominence of very high latitude, and which prevented the authors from properly evaluating the scattering angle acting upon the diagram.

Nevertheless, magnetic field vector determinations are always ambiguous, and these recent investigations do not address this problem. Two or more field vectors form the solution of the observed polarisation. The symmetries of scattering in the presence of a magnetic field were discussed in \citet{Bommier-80}. The first one is responsible for the same polarisation observed for two  field vectors that are  symmetrical with respect to the line-of-sight, in right-angle scattering. This symmetry is often referred to as fundamental ambiguity. When the scattering is not at a right angle, as in the present problem where we consider a prominence observation modified by solar rotation, the symmetry is modified also and the two ambiguous solutions are not symmetrical with respect to the line-of-sight. The two solutions may even correspond to slightly different field strengths. A second ambiguity is introduced by the Van Wleck ambiguity. In the present paper, we are not concerned by this ambiguity because we do not try to separate the two partially overlapping components of the \ion{He}{i} D$_3$ line and we analyse our data within the horizontal field hypothesis following the previously cited results. The Van Wleck ambiguity is not the same for two lines of different magnetic sensitivity. Thus, this ambiguity is automatically solved when analysing two such lines. A third ambiguity appears in the strong field\footnote{Note added in proof: in very strong field, two fields of opposite directions have the same effect.}, or saturated regime of the Hanle effect, which is for instance the case of the \ion{Fe}{xiv} 5303 \AA\, line and of the \ion{Fe}{xiii} 10,747 \AA\, and 10,798 \AA\, lines of the solar Corona. In the case of the saturated Hanle effect, the degeneracy is then of a factor $2^3 = 8$. 

Three methods able to solve the fundamental ambiguity have been proposed in the past. The first one, which we apply in the present paper, consists in comparing the two pairs of ambiguous solutions obtained for two different scattering angles, which are provided by observing the same quiescent prominence on two consecutive days. The symmetry is modified by solar rotation, and the comparison of the two pairs allows us to discriminate between the `true' solution, which is common to the two days, and the `mirror' or `ghost' solution, which is different for the two days. This method was investigated in \citet{Bommier-etal-81} and found to be successful, but the results were not given as they were thought to be too surprising at that time. Twenty prominences were analysed in this way. In most cases, the two ambiguous solutions are respectively directed on each side of the long axis of the filament. A filament is an elongated structure on the solar disk; its direction on the disk determines what is referred to as the long axis of the filament. This long axis lies along the photospheric neutral line below the prominence. As a result, in most cases the two ambiguous solutions each correspond to one of the two types of prominence magnetic structure: the Kippenhahn-Schl\"uter type, of `normal polarity', and the Kuperus-Raadu type, of `inverse polarity' \citep[see e.g.][]{Gibson-18}. The prominence polarity refers to the polarity, positive or negative, of the photospheric magnetic field on each side of the neutral line, which generally lies along the long axis of the filament. The magnetic field is horizontal in the prominence. When it crosses the neutral line from positive to negative, the magnetic model is of the Kippenhahn-Schl\"uter type and the polarity is said to be `normal'. When, on the contrary, the prominence magnetic field crosses the neutral line from negative to positive, the magnetic model is of the Kuperus-Raadu type and the polarity is said to be `inverse'. In the sample of twenty prominences, two were found to be of normal polarity and eighteen of inverse polarity. A large majority of inverse polarity prominences was then found.

Two other methods were later developed, both of which led to the same result: a large majority of inverse polarity prominences among the quiescent ones. \citet{Leroy-etal-84} developed a statistical analysis of the mirror symmetry of the two ambiguous solutions in a sample of  256 quiescent prominences, and derived the statistical result of 75\% inverse polarity prominences and 25\% normal polarity prominences. The normal polarity prominences were also found to be lower, sharp-edged, and with a stronger magnetic field. The third method involves the comparison of the two pairs of solutions provided by an optically thin line, like \ion{He}{i} D$_3$, and an optically thick line, like hydrogen H$\alpha$. When the prominence internal absorption and radiation cannot be ignored in the line formation model, in the optically thick case, the symmetry of the two ambiguous solutions is also modified. \citet{Bommier-etal-94} analysed such observations in fourteen quiescent prominences and found twelve of them to be of the inverse polarity type, and only two of them to be of the normal polarity type.

In the present paper we apply the first method for solving the ambiguity in a quiescent prominence observed with the TH\'EMIS telescope, by comparing the magnetic solutions obtained on two consecutive days. The previous ambiguity solutions of that type were achieved on Pic du Midi coronagraph data, where the observed polarisation was finally averaged over the whole object for accuracy purposes. The TH\'EMIS observations are accurate enough to avoid such an averaging and to obtain a field vector map of the prominence, as already published in the works cited above, but are ambiguous in those works. In Sect. \ref{sect--obs} we present the prominence observation, data treatment and Hanle inversion. We also detail our method for determining and propagating the ambiguity solution inside the prominence. We discuss the results in Sect. \ref{sect--res} and compare them with the magnetohydrodynamical (MHD) model of quiescent prominence by \citet{Aulanier-Demoulin-03} in Sect. \ref{conclusion}.

\section{Observation analysis}
\label{sect--obs}

\subsection{Observations}

\begin{figure}
\resizebox{\hsize}{!}{\includegraphics{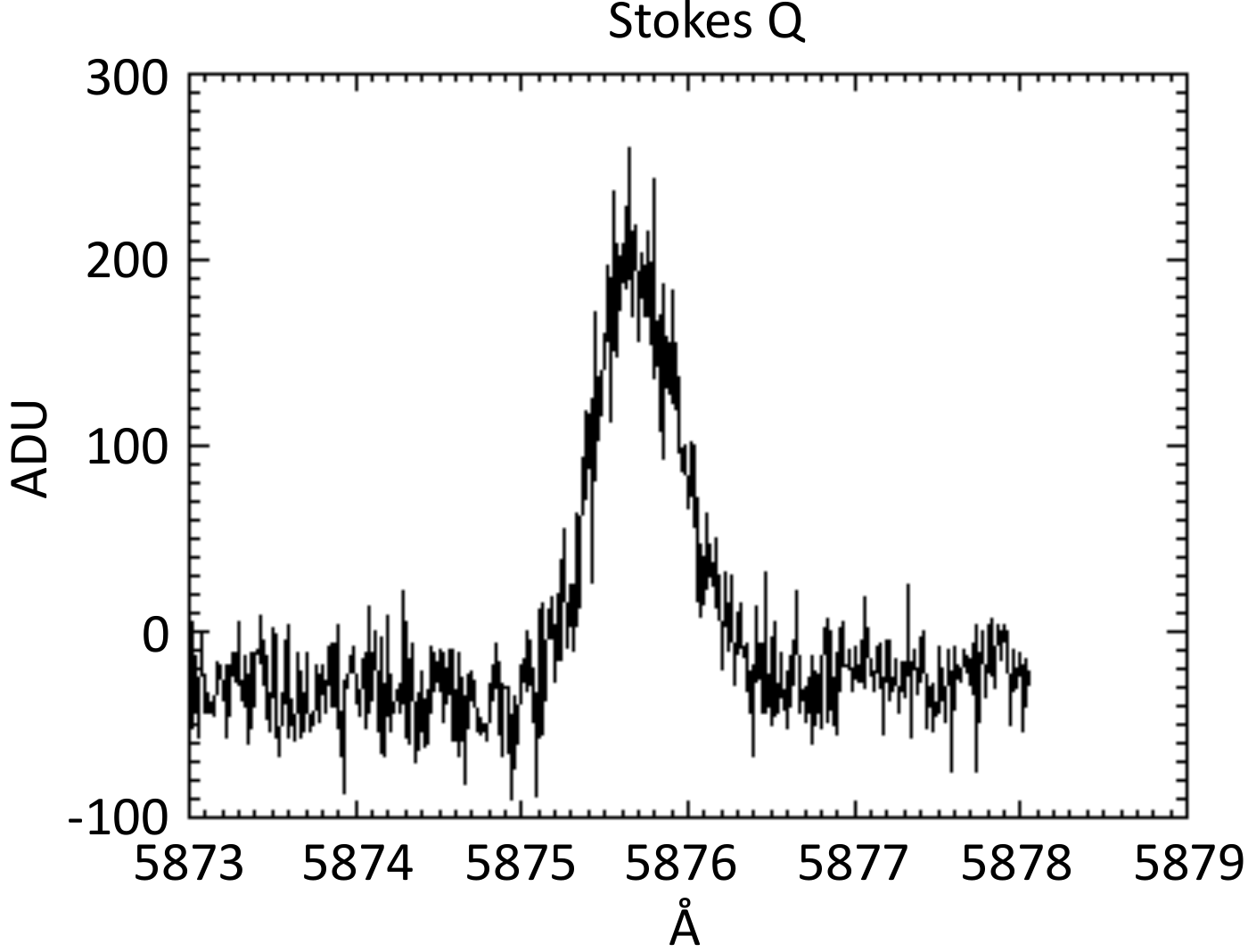}}
\caption{Typical Stokes $Q$ profile obtained in these observations.}
\label{StokesQ}
\end{figure}

\begin{figure}
\resizebox{\hsize}{!}{\includegraphics{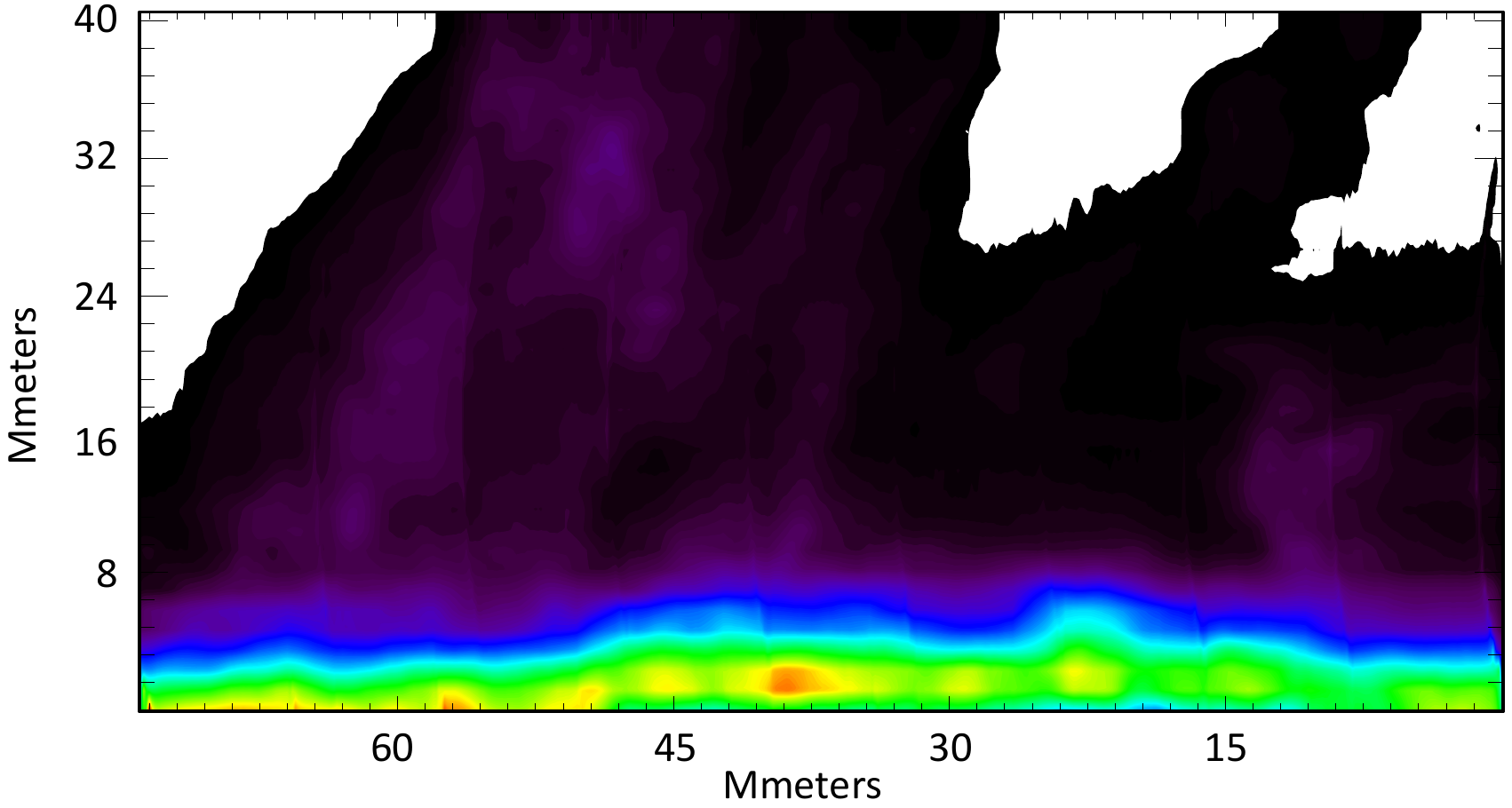}}
\resizebox{\hsize}{!}{\includegraphics{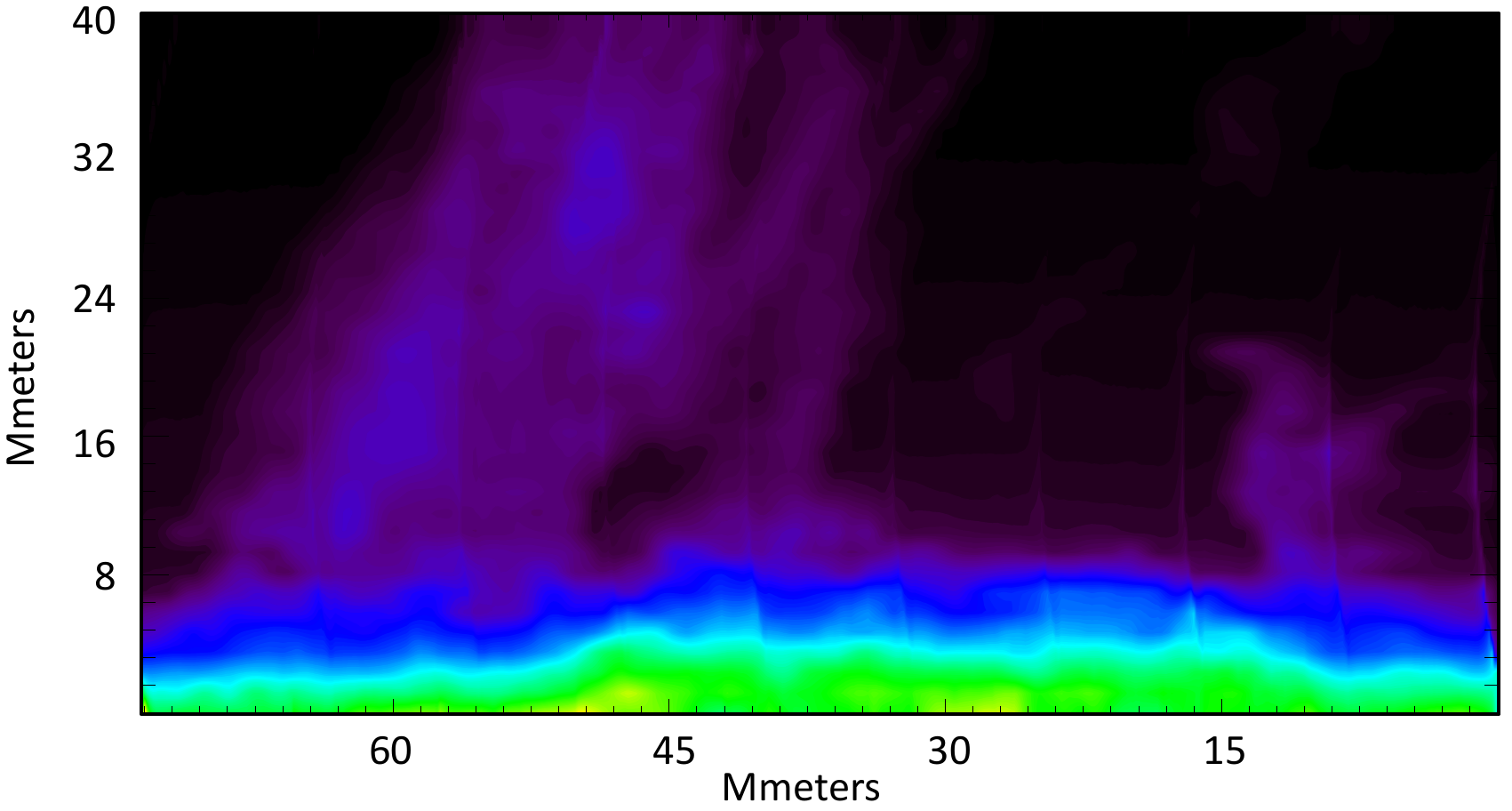}}
\caption{Prominence observed on 13 September 2008 at 11:48-13:40 UT at position angle 229$^{\circ}$. Upper image: Intensity at the centre of the \ion{He}{i} D$_3$ 5875.6 \AA\, line. Lower image: Intensity at the centre of the hydrogen H$\alpha$ line. The  bottom of the image is parallel to the solar limb. The warm colours (yellow, red) indicate the highest intensities, which are located at the solar limb.}
\label{StokesI-1309}
\end{figure}

\begin{figure}
\resizebox{\hsize}{!}{\includegraphics{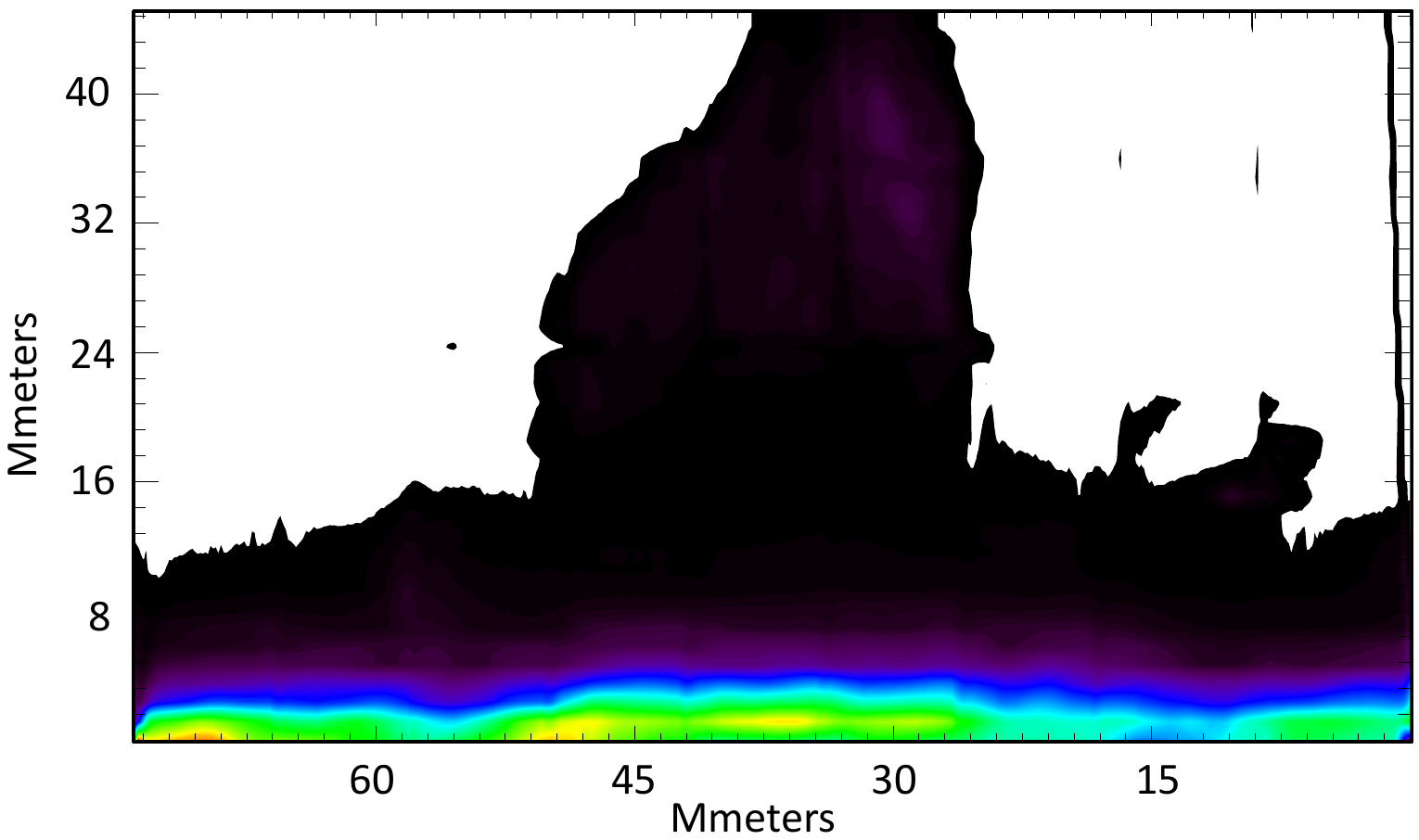}}
\resizebox{\hsize}{!}{\includegraphics{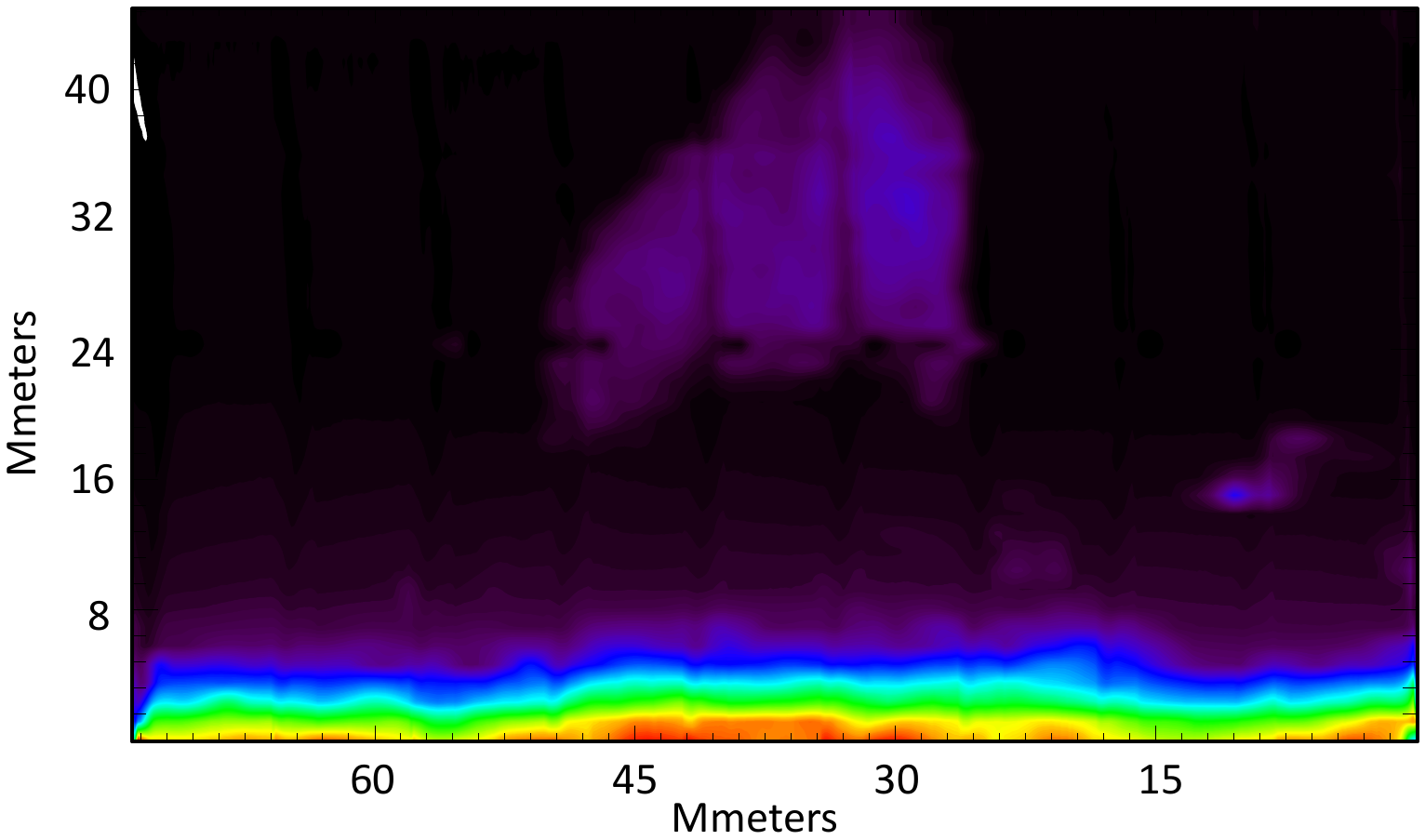}}
\caption{Prominence observed on 14 September 2008 at 11:19-13:34 UT at position angle 229$^{\circ}$. Upper image:  Intensity at the centre of the \ion{He}{i} D$_3$ 5875.6 \AA\, line. Lower image: Intensity at the centre of the hydrogen H$\alpha$ line. The  bottom of the image is parallel to the solar limb. The warm colours (yellow, red) indicate the highest intensities, which are located at the solar limb.}
\label{StokesI-1409}
\end{figure}

\begin{figure}
\resizebox{\hsize}{!}{\includegraphics{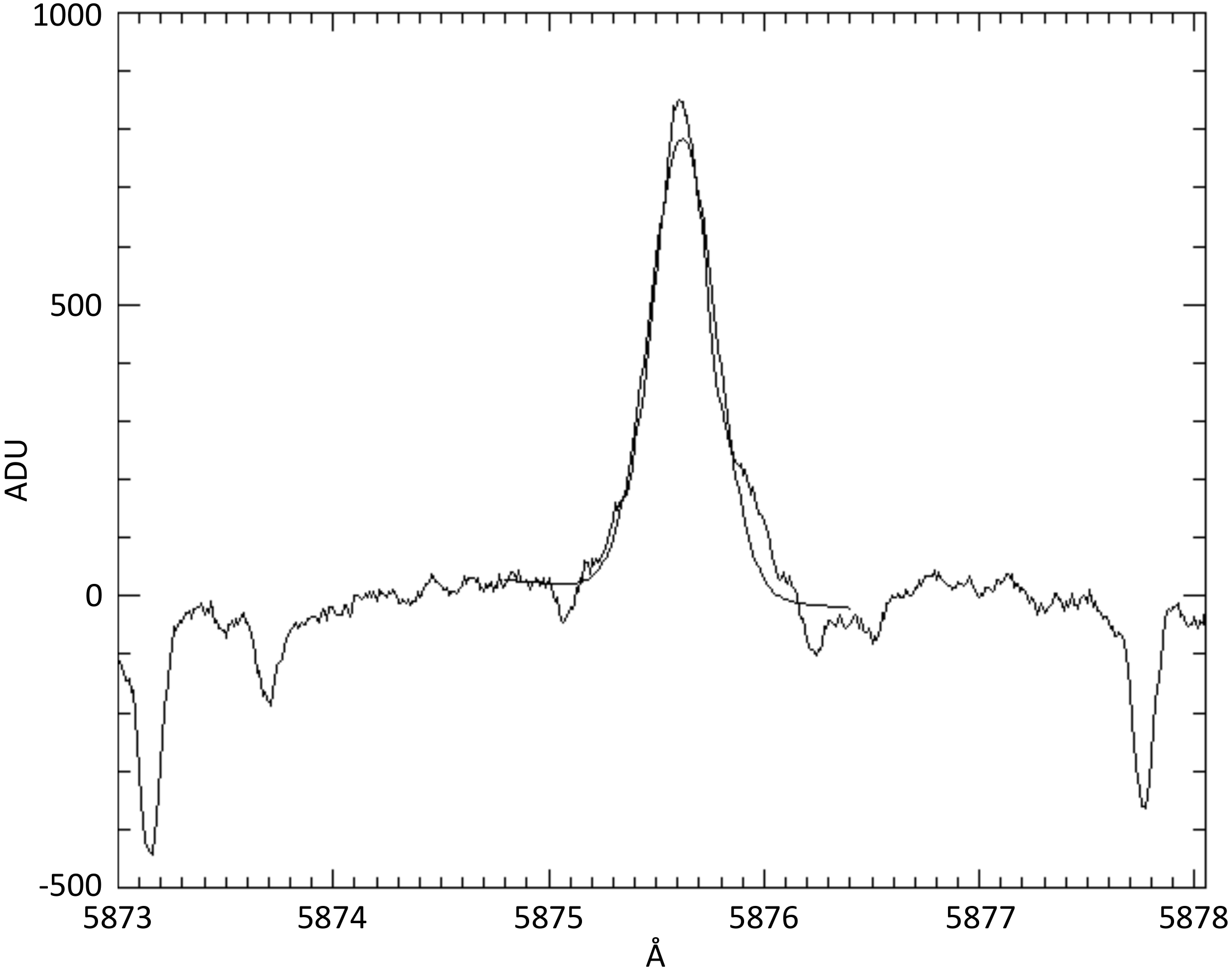}}
\resizebox{\hsize}{!}{\includegraphics{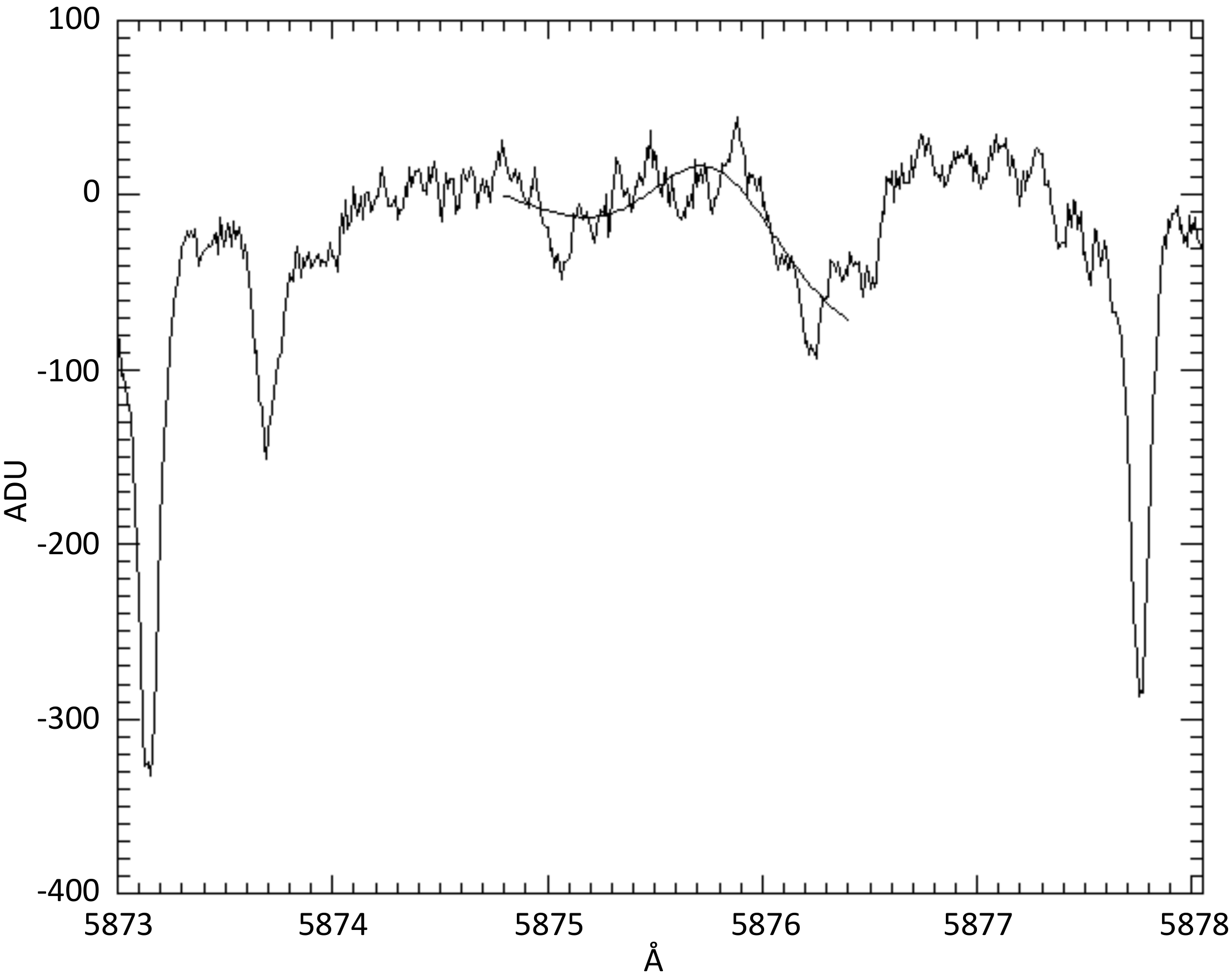}}
\caption{Example of a Gaussian fit of the \ion{He}{i} D$_3$ 5875.6 \AA\, line profile in our data. Top: Inside the prominence matter. Bottom: Outside the prominence matter. In the latter case,  the \ion{He}{i} D$_3$ line is visibly absent from the incident photospheric spectrum.}
\label{profil-fit}
\end{figure}

An observation campaign aimed at observing prominences on several consecutive days, in order to disambiguate the observed magnetic field vector by the Hanle effect, was led by V. Bommier at  the TH\'EMIS telescope from 11 to 17 September 2008. The prominence under study in this paper was observed on 13 September from 11:48 UT to 13:40 UT, and on 14 September from 11:19 UT to 13:34 UT, at the position angle 229$^{\circ}$.  On the H$\alpha$ spectroheliograms available in the BASS2000 database, we identified the corresponding filament in the preceding days before passing the limb and we verified that this filament or prominence lies far from active regions, hence its quiescent character. However, it is not a Polar Crown prominence, being located at rather low latitude. 

The TH\'EMIS telescope is polarisation-free, which means that the polarimetric analysis and beam-splitting are performed in F1 focus before any oblique reflection inside the telescope. Polarimetric calibration is therefore unnecessary. This polarimetric analysis is performed by splitting the beam into two beams whose respective intensities provide information about the radiation polarisation entering the telescope. Each polarisation Stokes parameter, $Q$, $U,$ or $V,$ is obtained by final subtraction of these two intensities, given the optical plate angular position in the beam-splitter, adapted to each $Q$, $U,$ or $V$ measurement. Three other positions are introduced to perform the beam-exchange for each of these Stokes parameters, that is, to exchange the beams in their respective ways inside the telescope. Such a beam-exchange technique notably increases the polarimetric accuracy of the measurement by suppressing or quantifying the effect of any other beam path difference \citep{Donati-etal-90}.

In the initial TH\'EMIS mode, the two separated beams were recorded by two different cameras, located at the ends of the two different beams. However, the two cameras cannot be strictly identically focused, and magnification differences between the two beams were  detected early on. In this respect, a `grid mode' was developed where the two beams are recorded on the same camera but after having passed through a mask that divides the image into three illuminated bands, each of 16.5 arcsec in width, separated by three analogous dark bands of the same size, into which the second beam bands are introduced in order to record the images of the two beams on the same camera. The bands are perpendicular to the spectrograph entrance slit. For these observations, TH\'EMIS was operated in this grid mode, which means that along the slit, there are three solar portions, each of 16.5 arcsec in length, alternated with the same portion but second beam.

Consequently, the three portions are separated by 16.5 arcsec in the solar image, or on the solar surface. Thus, a scan is performed along the slit with steps of 11 arcsec in order to scan the whole solar image or surface. Two  steps of 11 arcsec each are performed along the slit. However, the image on the camera is made of the radiation spectrum in the dimension perpendicular to the slit. The pixel size along the slit was 0.21 arcsec. Two lines were simultaneously observed: hydrogen H$\alpha$ and \ion{He}{i} D$_3$. The dispersion was 9.94 m\AA\ per pixel for \ion{He}{i} D$_3$ and 13.5  m\AA\ per pixel for H$\alpha$. A scan is then also performed perpendicular to the slit in order to reconstruct the solar image. The step of this scan was 1.5 arcsec. The scan along the slit was performed first, that is the two 11 arcsec steps along the slit were performed before moving the slit of 1.5 arcsec on the solar image. The slit was oriented parallel to the solar limb. The slit scan of 1.5 arcsec steps was performed along the solar radius from the limb towards the solar exterior. The duration of this 2D scan was 112 minutes on 13 September and 135 minutes on 14 September. Flat-field images were then acquired by rapidly and widely moving the slit at the Sun centre (with a shift from the centre on September 13 due to presence of activity at the centre of the Sun at that time). Camera dark current was finally recorded.

The polarimetric analysis code is mainly the one described in \citet{Bommier-Rayrole-02}, further extended to include beam-exchange by \citet{Bommier-Molodij-02}. The code performs dark current correction, destretching of flat field and raw spectro-images, and correction of the magnification difference between the two beams, even if the present application of the grid mode would make such a correction less necessary. Contrary to the first observations by \citet{Bommier-Rayrole-02}, the raw data are not finally averaged along the slit. A typical Stokes $Q$ profile thus obtained in these observations is displayed in Fig. \ref{StokesQ}.

The different bands of 16.5 arcsec width are then assembled to form the final image. These bands are partially overlapping. In the overlap, weights are assigned to each of the two overlapping bands for their combination. These weights are linearly varied from one band to the other, from one side to the other of the overlap. Figures \ref{StokesI-1309} and \ref{StokesI-1409} display the final image of the intensity at the centre of the line, for the two days and the two lines. These images were corrected for the scattered light by subtracting the neighbouring continuum intensity from the prominence line centre intensity. For the H$\alpha$ line, which contrary to the \ion{He}{i} D$_3$ line  is present as an absorption line in the incident photospheric spectrum, the neighboring continuum intensity was scaled by the ratio of the photospheric H$\alpha$ line centre intensity to its neighboring continuum.

The polarimetric accuracy of the observations was determined by the standard deviation of the intensity in a continuum region of the spectrum; it was found to be of about $6 \times 10^{-3}$ in each pixel, for \ion{He}{i} D$_3$. This has to be compared to the theoretically required polarimetric accuracy for measuring the magnetic field by interpretation of the Hanle effect, which we estimate to be $2 \times 10^{-3}$, given the typical \ion{He}{i} D$_3$ polarisation degree of 2-3\% in prominences \citep[see Table II of][with lower values in H$\alpha$]{Bommier-etal-94}. Thus, it is necessary to average over
at least ten pixels to reach the desired polarisation accuracy. Let us recall that our pixels are spatial in one dimension (along the slit) and spectral in the other dimension (perpendicular to the slit). The required averaging may therefore be spectral as well.

This is indeed the case for our observations. The \ion{He}{i} D$_3$ line in the prominence matter has the shape shown in the top panel of Fig. \ref{profil-fit}. We used information from a Gaussian fitting to localize the line, its maximum, and its half-width on each spectrum. The continuum was first subtracted by subtracting its linear fit. The line half-width was found to be about 30 spectral pixels. Thus, by summing over these 30 spectral pixels, the required polarisation accuracy is largely attained and any additional spatial averaging was not necessary for this purpose. The pixel size remained at 0.21 arcsec. We analysed line-integrated polarisations. Figure \ref{profil-fit} also shows how difficult it is to separate the two overlapping components of the \ion{He}{i} D$_3$ line in the analysis. In a first step, we did not investigate this question, and we integrate the whole profile as a single line, in our data as well as in our model.

Outside the prominence matter, the spectrum has the shape shown in the bottom panel of Fig. \ref{profil-fit},  where it can be seen that the \ion{He}{i} D$_3$ line is totally absent. This spectrum results from the incident photospheric radiation. The \ion{He}{i} D$_3$ line is clearly totally absent from the incident radiation. Therefore, any Doppler dimming or brightening is avoided in \ion{He}{i} D$_3$.

\subsection{Hanle inversion}
\label{subsect--Hanle}

The Hanle effect is the modification, due to the magnetic field, of the linear polarisation formed by scattering of anisotropic radiation. This is resonant scattering in a spectral line. In right-angle scattering, the scattered line becomes linearly polarised, even if the incident radiation is not. In right-angle scattering and magnetic field along the line of sight, the Hanle effect results in a depolarisation and a rotation of the linear polarisation direction about the magnetic field. In terms of Stokes parameters, these linear polarisation degree $p$ and polarisation direction referred to by the angle $\varphi$ it makes with the $Ox$ axis of the line-of-sight reference frame, are defined by
\begin{equation}
\begin{array}{l}
p = \frac{{\sqrt {{Q^2} + {U^2}} }}{I}\\
\cos 2\varphi  = \frac{Q}{{\sqrt {{Q^2} + {U^2}} }}\\
\sin 2\varphi  = \frac{U}{{\sqrt {{Q^2} + {U^2}} }}
\end{array}\ \ .
\end{equation}

The Hanle effect results from the Larmor rotation of the excited electron or atomic dipole about the magnetic field, combined with the finite lifetime of the line upper level. A synthetic presentation, with a table of line sensitivities as a function of their upper level lifetime, can be found in \citet{SahalB-81}. The effect is identical for all the Zeeman components of a line, and exists even if the Zeeman splitting is weak. Thus, the effect is wavelength independent. It is the same along the line profile, which can be integrated to increase accuracy, as we do in this paper. This is not spectropolarimetry. The first quantitative Hanle effect measurements in solar prominences \citep{Leroy-etal-77} were achieved through a filter, that is, without any spectral resolution.

In this respect, the first theories of the Hanle effect \citep{Bommier-77,Bommier-SahalB-78,Bommier-80} were developed without any line profile, because the effect is constant along it. The Hanle effect is due to the progressive destruction, as a function of the magnetic field, of the atomic so-called coherences, which are interference terms between different Zeeman sublevels, and which are represented by off-diagonal elements of the atomic density matrix. A complete theoretical explanation of the Hanle effect in a multilevel atom therefore requires the density matrix formalism to be applied, which was the case for the aforementioned initial theories.

The effect of the magnetic field on linear polarisation was presented in so-called diagrams, which are 2D representations. The linear polarisation is made of two parameters, namely $Q/I,U/I$ or $p,\varphi$, which are taken to define the two axes of the diagram. Examples of diagrams in $p,\varphi$ coordinates can be found in \citet[][see Figs. 5 and 6]{SahalB-etal-77}. An example of the Hanle effect diagram in $Q/I,U/I$ coordinates is the front cover illustration of the monograph by \citet{Landi-Landolfi-04}. The magnetic field has three coordinates, which are taken under the form of strength, inclination with respect to the solar radius, and azimuth with respect to the line-of-sight for limb observations. Thus, there is no bijection and a given linear polarisation represents not a single magnetic field but a series of possible magnetic fields. To reduce the number of variables, a diagram is plotted for each field inclination. A series of parametric curves are traced in the linear polarisation
coordinates, each representing the linear polarisation as a function of the magnetic field strength and for a given field azimuth angle. A second series of curves is plotted, each representing the linear polarisation as a function of the magnetic field azimuth angle and for a given field strength. These two series of curves intersect, forming irregular meshes. This is the so-called diagram mentioned above.

The inversion method consists in determining in what mesh lies the observed linear polarisation, which is represented by a point in these coordinates. Once the mesh is determined, a linear interpolation is achieved in it to determine the field strength and azimuth angle corresponding to the observed point. The mesh size was 5$^{\circ}$ in azimuth angle, and variable for the field strength because the Hanle effect is highly non-linear. This resolution method has to be applied for each field inclination. As a result, there is a series of possible magnetic field vectors compatible with an observed linear polarisation. This series can be represented as a function of the inclination angle. Examples are presented in \citet[][see e.g. Fig. 3]{Bommier-etal-81}.

The theoretical diagram is in fact computed for a given height above the limb, here 45 arcsec. The polarisation degree $p$ has to be normalised to the polarisation degree in zero magnetic field, which is also the maximum polarisation degree for exact right-angle scattering. We use $p_{\mathrm{max}}$ to denote this zero magnetic field polarisation degree, and the diagram coordinates are in fact $Q/I/p_{\mathrm{max}},U/I/p_{\mathrm{max}}$ or $p/p_{\mathrm{max}},\varphi$. Here $p_{\mathrm{max}}$ is a theoretical normalisation quantity, which has to be computed for each height above the solar limb, taking into account the true scattering angle of the observation, which may not be an exact right-angle. For this purpose, the solar coordinates of the prominence are determined by identifying it with a filament observed on the disk eight or so days before or after the prominence limb observation.

An extensive application of this inversion method can be found in \citet{Bommier-etal-81}. As explained in this publication, one single line is insufficient to determine the three components of the magnetic field vector from the two linear polarisation parameters. Is the circular polarisation profile the complementary information? This question was investigated  in \citet[][see Fig. 9]{Bommier-etal-81} and the conclusion was that it is not. The circular polarisation profile results from the Zeeman effect and provides the longitudinal magnetic field component, in weak fields as in prominences. Unfortunately, the weak field Zeeman effect is sensitive only to a field vector aligned to the line of sight, which is also the maximum sensitivity of the Hanle effect. The weak field Zeeman effect is insensitive to any transverse field, and the Hanle effect is insensitive to a vertical field (aligned with the solar radius), which is also transverse for limb observations. As for a field vector parallel to the limb, which is the other transverse direction, the Hanle effect is depolarisation only but no polarisation direction rotation occurs. The sensitivity of the Hanle effect  to such a field is therefore less than its sensitivity to a magnetic field aligned with the line of sight, where both depolarisation and polarisation direction rotation occur. This is the reason why knowledge of the longitudinal field component through the circular polarisation profile does not really bring additional information.

As explained above, the Stokes parameters have been integrated along the entire line profile. The \ion{He}{i} D$_3$ line is indeed made of six fine-structure components. Five of them are very close and grouped in the major component $3d^3D_{3,2,1} \rightarrow 2P^3P_{2,1}$. One lies apart as the minor component $3d^3D_1 \rightarrow 2P^3P_0$. The two components are separated by 0.34 \AA, which is not significantly different from the line width. The Hanle effect and the diagrams may be computed for each respective component. The upper level lifetimes are different. Thus, the Hanle sensitivity is different for these two components. These two components therefore form a line pair that is well-adapted to full magnetic field vector determination, by crossing the two series of solutions of the two components, as demonstrated in \citet{Bommier-etal-81}. The difficulty is in separating the two components in the observations in order to separately integrate over each component for further analysis. The top panel of Fig. \ref{profil-fit} shows their blend. In this paper, we took another approach and we integrated along the whole line profile, which we compared with the unresolved line diagram.

We based our analysis on the fact that the magnetic field is found to be mainly horizontal in quiescent prominences, as investigated by several authors and with different methods as stated in Sect. 1. Using this hypothesis of horizontal field, only two ambiguous field vectors form solutions of the observed integrated polarisation. These two solutions lie on either side of the long axis of the filament. The Carrington coordinates of the filament were found on the H$\alpha$ spectroheliograms available in the BASS2000 database. Our inversion code makes use of these coordinates to determine the true value of the scattering angle, which may not be a right-angle. The inclination angle of the solar rotation axis with respect to the sky plane, in other words the disk centre latitude, is also taken into account and was $b_0 = 7.22^{\circ}$ at the time of the observations. Finally, in order to determine the {direction of the long axis of the filament}, we better localized the filament channel in the EIT image at 195 \AA\  on 6 September 2008 at 05:36. In this image, we determined the solar azimuth 193.5$^{\circ}$ for this direction, with respect to the E--W oriented parallel direction.

From the polarimetric accuracy, our code is also able to derive the accuracy in field strength and direction determination in the diagrams by plotting the error box in the diagram coordinates. We thus obtained a mean inaccuracy of 0.3 G on the field strength and 1$^{\circ}$ on the field azimuth, due to polarimetric inaccuracy. We estimate the inaccuracy on the determination of the longitude and latitude of the filament on the solar surface to be  of 2$^{\circ}$, and we derived the related inaccuracy of 0.25 G on the field strength and 0.7$^{\circ}$ on the field azimuth. Finally, we estimate the inaccuracy on the solar limb position in our images to be  of
1 arcsec. This induces an inaccuracy of 0.15 G on the field strength and an inaccuracy of 0.3 $^{\circ}$ on the field azimuth. Accordingly, we estimate the global inaccuracy on the field strength  to be 0.7 G  and that on the field azimuth to be  2$^{\circ}$.

\subsection{Ambiguity resolution}

\begin{figure}
\resizebox{\hsize}{!}{\includegraphics{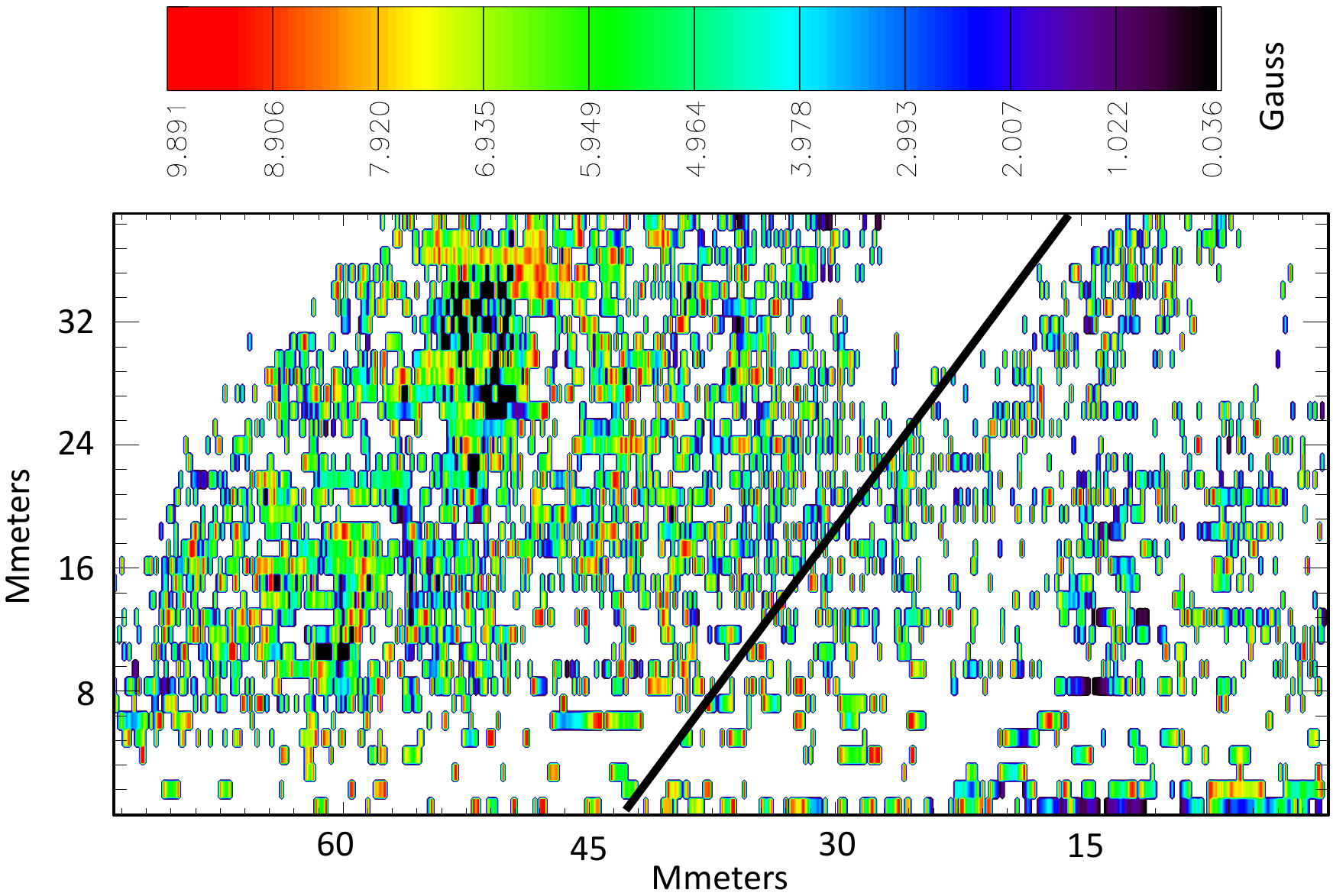}}
\resizebox{\hsize}{!}{\includegraphics{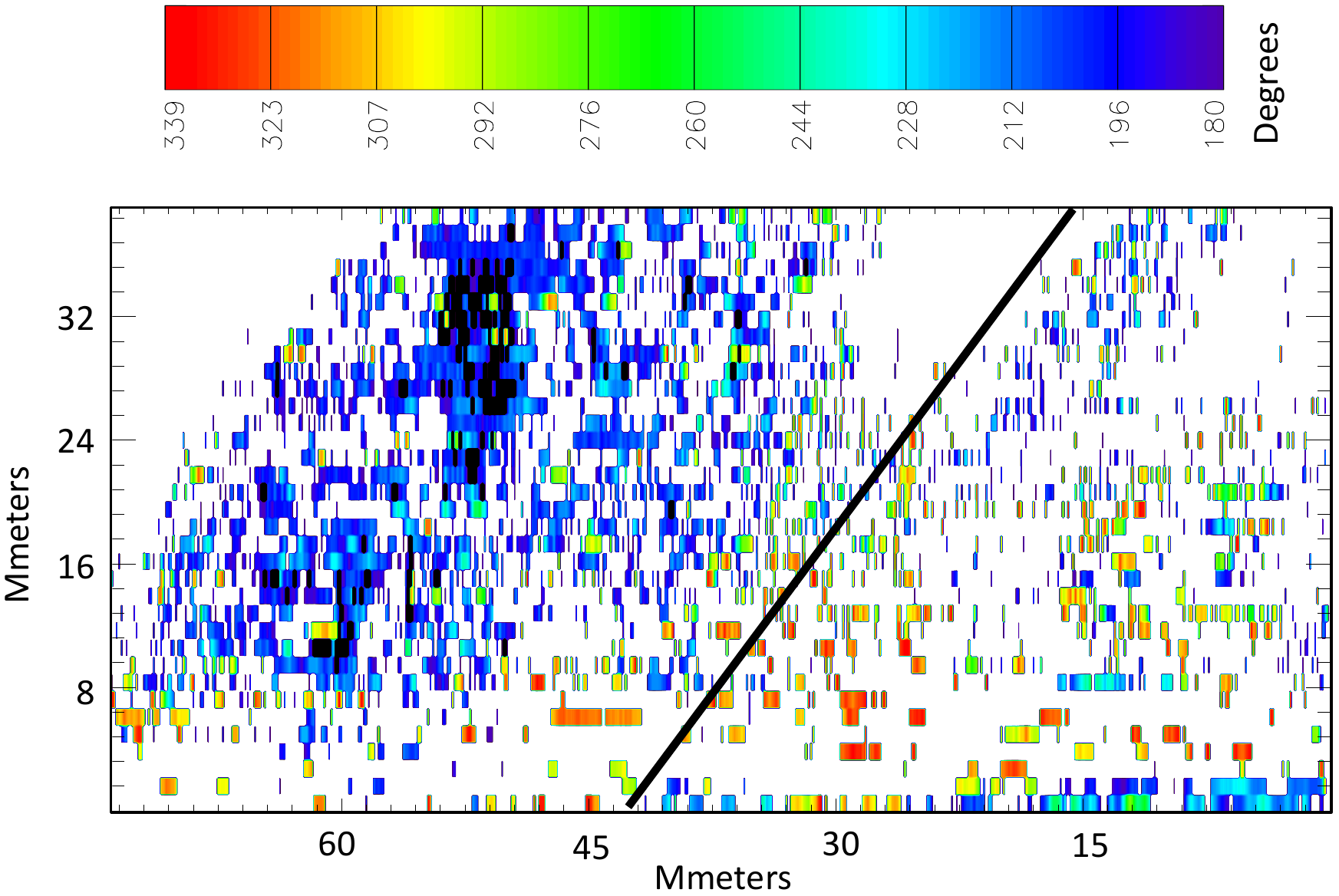}}
\caption{Field strength (top, in Gauss) and azimuth (bottom, in degrees counted with respect to the oriented E--W solar parallel) after ambiguity resolution in 4893 pixels. The magnetic field was determined within the horizontal field hypothesis. The  bottom of each image is parallel to the solar limb. The inclined black thick line delineates on the left-hand side the main body of the prominence for the gradient studies reported in Sect. \ref{sect--res}.}
\label{maps}
\end{figure}

\begin{figure}
\resizebox{\hsize}{!}{\includegraphics{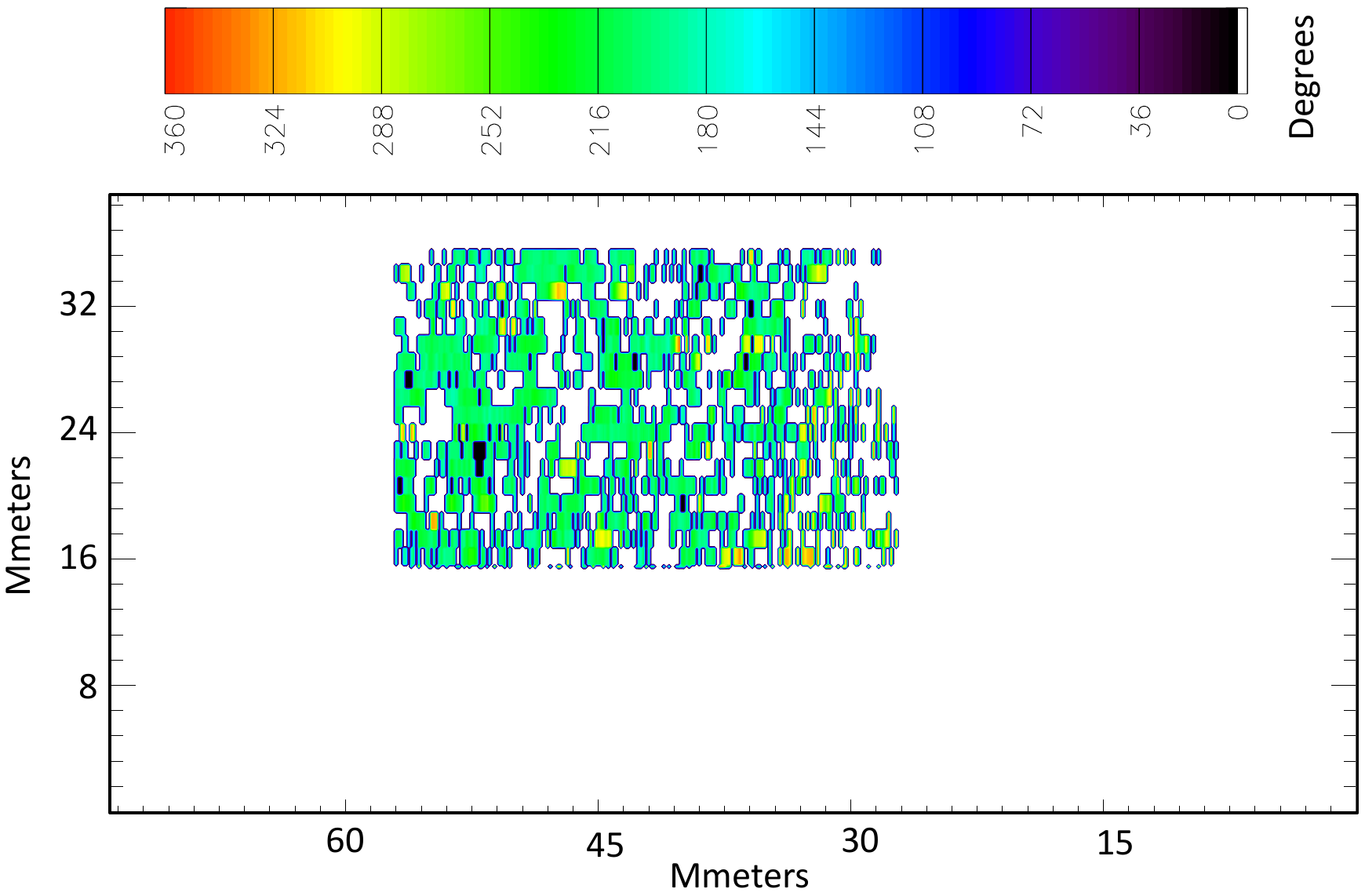}}
\caption{"Big pixel" in which the initial ambiguity resolution was performed by comparing the two  magnetic field solutions from consecutive days. The image frame is the same as in Fig. \ref{StokesI-1309} and in Fig. \ref{maps}. The quantity represented in colour is the azimuth after ambiguity resolution, in each small pixel, in degrees counted with respect to the oriented E--W solar parallel.}
\label{bigpixel}
\end{figure}

The ambiguity was first solved by comparing the two pairs of solutions on the two consecutive days, in a large averaged portion of the prominence, the "big pixel". This is represented in Fig. \ref{bigpixel}, the frame of which is the same as that of Fig. \ref{maps}. We averaged the observed Stokes parameters in the whole big pixel and determined from them the average magnetic field vector for the two consecutive days following the method described in Sect. \ref{subsect--Hanle}. The solution was then propagated from pixel to neighbouring pixel, from the prominence centre to the prominence boundary, following the closest-neighbour technique as in the AZAM code \citep{Lites-etal-95}. The propagation algorithm is as follows: once the ambiguity is solved in a given pixel, leading to a single field vector solution in this pixel, this vector is compared to the two vector solutions of the neighbouring pixel. The retained solution is the one that makes the smallest angle with the resolved pixel solution. This method is referred to as the acute angle method.

Initially, our two images contained 17290 pixels. For 2546 of them, no or a very faint \ion{He}{i} D$_3$ line was visible as shown in the bottom of Fig. \ref{profil-fit}. For 4615 pixels, the Hanle inversion provided no solution. The magnetic field may not be horizontal, or this is an effect of the above listed inaccuracies. For about 500 pixels only, the two ambiguous field vectors were surprisingly found to lie on the same side of the line of sight. We discarded these few pixels from our analysis. 

We first averaged the Stokes parameters, integrated along the line profile, in 3903 (13 September) and 3187 (14 September) pixels representing the "big pixel". We then performed the Hanle inversion on these averaged Stokes parameters. The azimuths of the two solutions were found as 131.55$^{\circ}$ and 207.28$^{\circ}$ on 13 September, and 144.45$^{\circ}$ and 215.21$^{\circ}$ on 14 September, leading to differences of 12.91$^{\circ}$ and 7.92$^{\circ}$, respectively. The difference between these latter two differences is larger than our angular inaccuracy discussed above. The `true' solution is therefore the second one.

We then propagate this solution across the prominence. Due to the absence of the spectral line or solutions, the eliminated pixels made the propagation difficult at times. Finally, we solved the ambiguity in 4893 pixels of each image. The results are presented in Fig. \ref{maps}, where the white zones correspond to eliminated pixels.

\section{Results and Discussion}
\label{sect--res}

\begin{figure}
\resizebox{\hsize}{!}{\includegraphics{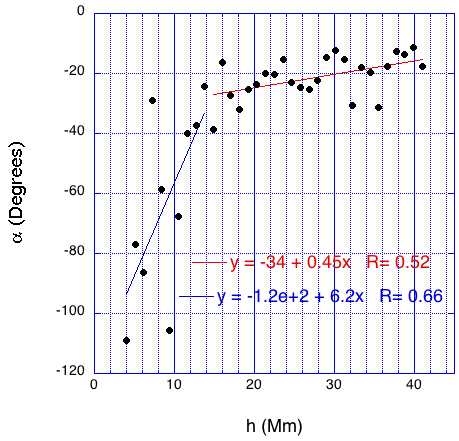}}
\caption{Average angle between the horizontal magnetic field vector and the long axis of the filament, as a function of the height in the prominence. The field has been averaged at each height. The minus sign of the angle indicates the inverse polarity, with respect to the neighbouring photospheric polarities.}
\label{alpha-h}
\end{figure}

\begin{figure}
\resizebox{\hsize}{!}{\includegraphics{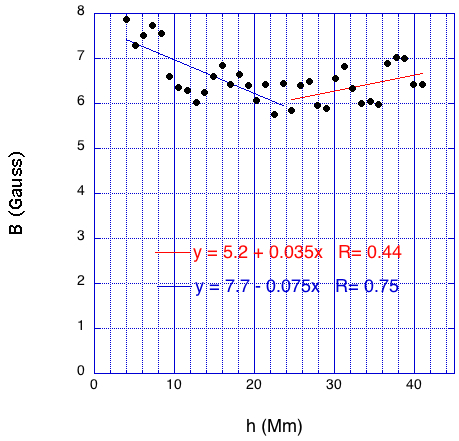}}
\caption{Average magnetic field strength as a function of the height in the prominence. The field has been averaged at each height.}
\label{B-h}
\end{figure}

In the following, we present our results as averages at each height in the prominence. Only the prominence body, defined as the region at the left-hand side of the inclined black thick line of Fig. \ref{maps}, was retained for these averages.

Figure \ref{alpha-h} presents the variation with height of the $\alpha$ angle, which is the angle between the horizontal field vector and the long axis of the filament. As the long axis of the filament lies along the photospheric magnetic neutral line, which separates two regions of opposite polarity, the sign of the $\alpha$ angle was assessed following the prominence field polarity with respect to these neighbouring photospheric polarities. A negative $\alpha$, as we obtain, means that the prominence is of the inverse polarity type. 

Two regions can be distinguished in our plot of Fig. \ref{alpha-h}: above and below 15 Mm. Above 15 Mm, we obtain an average value of $\alpha = -21^{\circ}$, in excellent agreement with the $\alpha = -25^{\circ}$ obtained by \citet{Leroy-etal-84} for 75\% of the prominences found to be of the inverse polarity type. In another paper devoted to polar crown prominences, where this angle can be determined without an ambiguity solution when the prominences are seen edge-on, \citet{Leroy-etal-83} obtain the same average value of  $\alpha = -25^{\circ}$. With the spatial resolution in the TH\'EMIS prominence, we obtain that this angle decreases in absolute value with increasing height, at a rate of $-4.5 \times 10^{-4}$ degrees/km.

The lower part of the plot displays a faster decrease of $-6 \times 10^{-3}$ degrees/km. Such very low distances with respect to the solar limb were not accessible at the time of previous measurements, which were made with coronagraphs.

Figure \ref{B-h} displays the variation in field strength. We obtain a typical field strength of 6 G  for prominences. In the upper part, we obtain a gradient of field strength of $0.35 \times 10^{-4}$ G/km, in good agreement with the value of $0.5 \times 10^{-4}$ G/km obtained by \citet{Leroy-etal-83}. 

On the contrary, the lower part of the plot displays a decrease in the field strengths, in those low altitudes which were inaccessible to the coronagraphs. The decreasing rate we find is $-0.75 \times 10^{-4}$ G/km. The limit between the two behaviours is 25 Mm, higher for the field strengths than for the field azimuth, where this limit is 15 Mm.

\section{Conclusion}
\label{conclusion}

In the upper part (higher than 15 Mm) of this prominence observed with TH\'EMIS, we obtain an average field strength of 6.4 G and an average angle of $-21^{\circ}$ between the field (assumed to be horizontal) and the long axis of the filament, in the inverse polarity scheme. This is in excellent agreement with the previous measurements listed in Sect. 1. The new result concerns the decrease in the absolute value of this angle with increasing height. We obtain a gradient of $-4.5 \times 10^{-4}$ degrees/km for the absolute value of this angle, in the inverse polarity scheme for the prominence magnetic field (see Fig. \ref{alpha-h}). For the field strength, we obtain a positive gradient of $0.35 \times 10^{-4}$ G/km, in very good agreement with the previous determination by \citet{Leroy-etal-83} (see Fig. \ref{B-h}).

This new result regarding the gradient of the horizontal field azimuth with respect to the long axis of the filament is in excellent agreement with the MHD model of a quiescent filament by \citet[][see their Fig. 2]{Aulanier-Demoulin-03}. These latter authors also report a decrease in the absolute value of this angle with increasing height in the upper part of the prominence.

With our TH\'EMIS observations, we are also able to determine the field at low altitudes in the prominence, inaccessible to the previous coronagraphic observations. For the field azimuth with respect to the long axis of the filament, we obtain a faster decrease with height of the absolute value of this angle, of $-6 \times 10^{-3}$ degrees/km. Figure 2 of \citet{Aulanier-Demoulin-03} displays more scattered values of this angle at lower altitudes, but the general trend is also a faster decrease with height of the absolute value of this angle, in agreement with our observations. 

As for the field strength, \citet{Aulanier-Demoulin-03} find increasing field strength with height also at low altitudes; this increase is even faster than at higher altitudes. On the contrary, we obtain an average decrease of $-0.75 \times 10^{-4}$ G/km in the lower regions of the prominence. In their paper, \citet{Aulanier-Demoulin-03} assess that the highest field strength values always lie at low altitude, as visible in their Fig. 2. Their obtained average gradient of the field strength with altitude is however positive. They point out that at low altitudes, the magnetic field is made of a mixing of independent structures, which may be complex, with not necessarily high \ion{He}{i} D$_3$ emission. In our data, a larger number of pixels were eliminated in the low-altitude region due to failure in the inversion or to absence of \ion{He}{i} D$_3$ emission. Our data are therefore dominated by more intense \ion{He}{i} D$_3$ emission, which may have biased the field strength we observe with respect to the general field strength in this region. In addition, it is difficult to precisely locate the solar limb, which prevents us from obtaining precise results at very low altitudes.

Our results, in terms of field strength, are significantly different from those of \citet{Kuridze-etal-19}, because the prominence we are studying is a quiescent one, located far from any active region and/or out of any eruptive phase. On the contrary, \citet{Kuridze-etal-19}  studied coronal flare loops, which were located above an active region and observed during an eruption, even if a prominence eruption may be different from a solar flare. In a general manner, it would certainly be fruitful to complement vector magnetic field observations with vector velocity field observations, as \citet{Kuridze-etal-19} do.

\begin{acknowledgements}
The TH\'EMIS observations were supported by the Programme National Soleil-Terre (PNST) of the French Centre National de la Recherche Scientifique (CNRS).
\end{acknowledgements}

\bibliographystyle{aa}
\bibliography{bommierrefs}

\end{document}